\documentclass{jpsj3}
\usepackage{txfonts}
\usepackage{overcite,mathptm,times,graphicx}
\usepackage{subfigure}
\newcounter{theorem}
\renewcommand\thetheorem{\arabic{section}.\arabic{theorem}}

\newenvironment{proof}{\par\noindent{\bf Proof.}
}{\proofbox\par\medskip}

\newdimen\LENB \newdimen\LENW \newdimen\THI
\newdimen\LENWH \newdimen\LENTOT \newcount\N
\def\vbrknlnele#1#2#3{
  \LENB=#1pt \LENW=#2pt \THI=#3pt
  \LENWH=\LENW \divide\LENWH by 2
  \LENTOT=\LENB \advance\LENTOT by \LENW
  \vbox to \LENTOT{
    \vbox to \LENWH{}
    \nointerlineskip
    \vbox to \LENB{\hbox to \THI{\vrule width \THI height \LENB}}
    \nointerlineskip
    \vbox to \LENWH{}
  }}

\def\vbrknln#1{
  \N=#1
  \vcenter{
    \vbox{
      \loop\ifnum\N>0
        \vbox to 4pt{\vbrknlnele{2}{2}{0.1}}
        \nointerlineskip
        \advance\N by -1
      \repeat
  }}}

\def\hbrknlnele#1#2#3{
  \LENB=#1pt \LENW=#2pt \THI=#3pt
  \LENTOT=\LENB \advance\LENTOT by \LENW
  \vcenter{
    \vbox to \THI{
      \hbox to \LENTOT{
        \hfil
        \vrule width \LENB height \THI
        \hfil}
  }}}

\title{General Mixed Multi-Soliton Solutions to One-Dimensional Multicomponent Yajima-Oikawa System}

\author{Junchao Chen$^{1,2}$, Yong Chen$^1$\thanks{ychen@sei.ecnu.edu.cn}, Bao-Feng Feng$^2$\thanks{feng@utpa.edu}, Ken-ichi Maruno$^{3}$\thanks{kmaruno@waseda.jp}}
\inst{$^1$Shanghai Key Laboratory of Trustworthy Computing, East China Normal University, Shanghai, 200062, People's Republic of China \\
$^2$ Department of Mathematics, The University of Texas-Pan American, Edinburg, TX 78541, USA \\
$^3$~Department of Applied Mathematics, School of Fundamental Science
and Engineering, Faculty of Science and Engineering,
Waseda University, 3-4-1 Okubo, Shinjuku-ku, Tokyo 169-8555, Japan} %\\

\abst{
In this paper, we derive a general mixed (bright-dark) multi-soliton solution to a one-dimensional multicomponent Yajima-Oikawa (YO) system, i.e., the ($M+1$)-component YO system comprised of $M$-component short waves (SWs) and one-component long wave (LW)
for all possible combinations of nonlinearity coefficients including positive, negative and mixed types.
With the help of the KP-hierarchy reduction method,
we firstly construct two types of general mixed $N$-soliton solution (two-bright-one-dark soliton and one-bright-two-dark one for SW components) to the (3+1)-component YO system in detail.
Then by extending the corresponding analysis to the ($M+1$)-component YO system, a general mixed $N$-soliton solution in
Gram determinant form is obtained.
The expression of the mixed soliton solution also contains the general all bright and all dark $N$-soliton solution as special cases.
Besides, the dynamical analysis shows that the inelastic collision can only take place among SW components when at least two SW components have bright solitons in mixed type soliton solution.
Whereas, the dark solitons in SW components and the bright soliton in LW component always undergo usual elastic collision.}

\begin{document}
\maketitle

\vspace{5cm}

\section{Introduction}

The investigation of multicomponent nonlinear systems
has received much attention in recent years \cite{hasegawa1995solitons,kivshar2003optical,ablowitz2004discrete,manakov1974theory}.
Of particular concern is the multicomponent
generalization of the nonlinear Schr\"{o}dinger (NLS) equation, namely, the vector NLS equation \cite{ablowitz2004discrete,manakov1974theory,hisakado1995integrable}.
Particularly, it has been shown that multicomponent bright solitons undergo shape
changing collisions with intensity redistribution  \cite{radhakrishnan1997inelastic,radhakrishnan1999coupled,kanna2001exact,kanna2006soliton,vijayajayanthi2008bright,vijayajayanthi2009multisoliton}.
This interesting collision feature has been widely studied
in real physical systems such as nonlinear optics and Bose-Einstein condensates \cite{sukhorukov2003multiport,babarro2005controllable,jakubowski1998state,steiglitz2001time,kanna2003exact}.

The long-wave-short-wave resonance interaction (LSRI) is a fascinating physical process,
in which a resonant interaction takes
place between a weakly dispersive long-wave (LW) and a short-wave (SW)
packet when the phase velocity of the former exactly
or almost matches the group velocity of the latter.
The theoretical investigation of this LSRI was first done by Zakharov \cite{zakharov1972collapse} on Langmuir waves in plasma.
In the case of long wave propagating in one direction,
the general Zakharov system was reduced to the one-dimensional (1D) Yajima-Oikawa (YO) equation \cite{yajima1976formation}.
This model equation also appears in diverse areas that include hydrodynamics \cite{benny1977general}, nonlinear optics \cite{kivshar1992stable,chowdhury2008long}, biophysics \cite{davydov1991solitons} etc.
The 1D YO system is integrable by the inverse scattering transform method \cite{yajima1976formation} and
 admits both bright and dark soliton solutions \cite{ma1979some,ma1978complete}.
The rogue wave solutions to the 1D YO system have
recently been derived by using the Hirota's bilinear method
\cite{wing2013rogue} and Darboux transformation \cite{chen2014dark,chen2014darboux}.

In the present paper, we consider a general multicomponent 1D YO system \cite{chen2014multidark}
\begin{equation}\label{mix1d-01}
\begin{array}{l}
\textmd{i}S^{(\ell)}_t  - S^{(\ell)}_{xx}+ L S^{(\ell)}=0,\ \ \ell =1,2,\cdots, M,\ \ \ \ \ \\
L_t=2\sum^M_{\ell=1}\sigma_\ell|S^{(\ell)}|^2_x, \ \ \sigma_\ell=\pm1,
\end{array}
\end{equation}
which describes the dynamics of nonlinear resonant interaction between a LW ($L$) and
multiple (say $M$) SWs ($S^{(\ell)}$) in one-dimensional case.
Hereafter, we refer to the above multicomponent system as the ($M$+1)-component YO system.
It was pointed out that the (2+1)-component YO system
can be deduced from a set of three coupled NLS equations governing the propagation of three optical fields in a triple mode optical fiber by applying the asymptotic reduction procedure \cite{kanna2013general}.
Eqs.(\ref{mix1d-01}) has been derived to describe the interaction between a quasi-resonance circularly polarized optical pulse and a long-wave electromagnetic spike \cite{sazonov2011vector}.
In the context of the many-component magnon-phonon system, such multicomponent YO-type system has also been proposed and
its corresponding Hamiltonian formalism was studied \cite{myrzakulov1986particle}.
Also, the authors in ref.\cite{kanna2013general} have carried out Painlev\'{e} analysis for Eqs.(\ref{mix1d-01}) and  obtained the general
bright N-soliton solution in the Gram determinant form.
Later on, they constructed an extensive set of exact periodic solutions in terms of Lam\'{e} polynomials of order one and two \cite{khare2014elliptic}.
The rogue waves of the (2+1)-component YO system with $\sigma_1=\sigma_2=1$ have been reported in Ref.\cite{chen2014coexisting}.

It is worth mentioning that the KP-hierarchy reduction method to derive soliton equations as well as their solutions was developed by the Kyoto school \cite{jimbo1983solitons}.
A number of soliton equations such as the NLS equation, the modified KdV equation,
and the Davey--Stewartson equation can be reduced from the KP-hierarchy.
Indeed, the multicomponent YO system (\ref{mix1d-01})
with same nonlinearity coefficients $\sigma_\ell$ (all $\sigma_l=1$,or $\sigma_l=-1$) was derived from the KP-hierarchy in \cite{zhang1994solutions,sidorenko1993multicomponent,liu1996bi}.
In particular, general (pseudo-) reductions of the two-dimensional Toda lattice hierarchy to constrained KP systems with dark soliton solutions were introduced in \cite{willox1999an}  and reductions to constrained KP systems with bright soliton solutions from the multi-component KP hierarchy were introduced in \cite{willox1999kp}.
For the (1+1)-component YO system (when $M=1$ in  Eqs.(\ref{mix1d-01})),
the detailed interpretation of how to obtain the Wronski-type bright and dark soliton solutions by using the reduction technique
was presented in Ref.\cite{willox1999pseudo}.
By applying this method, general dark-dark soliton solution was derived in a coupled focusing--defocusing NLS equation \cite{ohta2011general}.
Recently, one of the authors \cite{feng2014general} has constructed general bright-dark $N$-soliton solution to the vector
NLS equation of all possible combinations of nonlinearities including all focusing,
all-defocusing and mixed types.
In a previous study \cite{chen2014multidark}, we have obtained both the Gram-type and Wronski-type determinant solutions of $N$ dark-soliton for Eqs. (\ref{mix1d-01}).

The goal of this paper is to construct general bright-dark multi-soliton solution to the multicomponent YO system (\ref{mix1d-01}).
The rest of the paper is arranged as follows.
In section 2, we deduce two types of general mixed soliton solution, which include two-bright-one-dark and one-bright-two-dark soliton for SW components, to the (3+1)-component YO system by using the KP-hierarchy reduction technique.
In section 3, general bright-dark soliton solution consisting of $m$ bright solitons and $M-m$ dark
solitons to the multicomponent YO system (1) is obtained by generalizing our analysis.
We summarize the paper in section 4.

\section{General mixed-type soliton solution to the one-dimensional (3+1)-component YO system}

Prior to presenting the general mixed-type soliton solution for Eq.(\ref{mix1d-01}), we first consider the 1D (3+1)-component YO system
\begin{equation}\label{mix1d-02}
\begin{array}{l}
\textmd{i}S^{(1)}_t  - S^{(1)}_{xx}+ L S^{(1)}=0,\ \  \\
\textmd{i}S^{(2)}_t  - S^{(2)}_{xx}+ L S^{(2)}=0,\ \  \\
\textmd{i}S^{(3)}_t  - S^{(3)}_{xx}+ L S^{(3)}=0,\ \ \\
L_t=2(\sigma_1|S^{(1)}|^2+\sigma_2|S^{(2)}|^2+\sigma_3|S^{(3)}|^2)_x,
\end{array}
\end{equation}
where $\sigma_\ell=\pm1$ for $\ell=1,2,3$.
For three short-wave components, there exist two kinds of mixed-type vector solitons
(two-bright-one-dark and one-bright-two-dark).
Hence, in the subsequent two subsections, we will construct these two types of soliton solutions,
respectively.

\subsection{Two-bright-one-dark soliton for the SW components}

In this case, we assume that the SW components $S^{(1)}$ and $S^{(2)}$ are of bright type and the SW component $S^{(3)}$ is of dark type.
By introducing the dependent variable transformations
\begin{equation}\label{mix1d-03}
\hspace{-1cm} S^{(1)}=\frac{g^{(1)}}{f},\ \ S^{(2)}=\frac{g^{(2)}}{f},
\ \ S^{(3)}=\rho_1\frac{h^{(1)}}{f} {\rm e}^{{\rm i}(\alpha_1 x+ \alpha^2_1t)},
\ \ L=-2(\log f)_{xx},\ \
\end{equation}
equations.(\ref{mix1d-02}) are converted into the following bilinear
equations:
\begin{equation}\label{mix1d-04}
\begin{array}{l}
\left[ \textmd{i} D_t-D^2_x \right]g^{(k)}\cdot f=0,\ \ k=1,2,\\
\left[ \textmd{i} (D_t-2\alpha_1D_x)-D^2_x \right]h^{(1)}\cdot f=0,\ \ \\
\left(D_tD_x-2\sigma_3\rho^2_1\right)f\cdot f= -2\sum^2_{k=1}\sigma_k g^{(k)}g^{(k)*}-2\sigma_3\rho^2_1 h^{(1)}h^{(1)*},
\end{array}
\end{equation}
where $g^{(1)},g^{(2)}$ and $h^{(1)}$ are complex-valued functions, $f$ is a real-valued function,
$\alpha_1$ and $\rho_1$$(\rho_1>0)$ are arbitrary real constants and ${}^*$ denotes the complex
conjugation hereafter.
The Hirota's bilinear differential operator is defined by \cite{hirota2004direct}
\begin{eqnarray*}
D^n_xD^m_t(a\cdot b)=\bigg( \frac{\partial}{\partial x} - \frac{\partial}{\partial x'} \bigg)^n
\bigg( \frac{\partial}{\partial t} - \frac{\partial}{\partial t'} \bigg)^m
a(x,t)b(x',t')\bigg|_{x=x',t=t'} .
\end{eqnarray*}

In what follow, we will show how the (3+1)-component YO system and its mixed type multisoltion solution can be
obtained from three-component KP hierarchy by suitable reductions.
To this end, we firstly present tau functions of the Gram determinant form in three-component KP hierarchy
\begin{eqnarray}
\label{mix1d-05}&& \tau_{0,0}({k_1}) =
\left|
\begin{array}{ccc}
A & I \\
-I & B
\end{array}
\right|,\\
\label{mix1d-06}&& \tau_{1,0}(k_1)=
\left|
\begin{array}{ccc}
A & I & \Phi^T \\
-I & B & \textbf{0}^T\\
\textbf{0} & -\bar{\Psi} & 0
\end{array}
\right|,\ \
 \tau_{-1,0}(k_1)=
\left|
\begin{array}{ccc}
A & I & \textbf{0}^T \\
-I & B & \Psi^T\\
-\bar{\Phi} & \textbf{0} & 0
\end{array}
\right|,\\
%%%%%%%%%%%%%%%%%%
\label{mix1d-07}&& \tau_{0,1}(k_1)=
\left|
\begin{array}{ccc}
A & I & \Phi^T \\
-I & B & \textbf{0}^T\\
\textbf{0} & -\bar{\Upsilon} & 0
\end{array}
\right|,\ \
 \tau_{0,-1}(k_1)=
\left|
\begin{array}{ccc}
A & I & \textbf{0}^T \\
-I & B & \Upsilon^T\\
-\bar{\Phi} & \textbf{0} & 0
\end{array}
\right|,
\end{eqnarray}
where $I$ is an $N\times N$ identity matrix, $A$ and $B$ are $N\times N$ matrices whose entries are
\begin{eqnarray*}
\hspace{-1cm} a_{ij}(k_1)=\frac{1}{p_i+\bar{p}_j} \left(-\frac{p_i-c_1}{\bar{p}_j+c_1} \right)^{k_1} {\rm e}^{\xi_i+\bar{\xi}_j},\ \
b_{ij}=\frac{1}{q_i+\bar{q}_j} {\rm e}^{\eta_i+\bar{\eta}_j} + \frac{1}{r_i+\bar{r}_j} {\rm e}^{\chi_i+\bar{\chi}_j},
\end{eqnarray*}
and $\textbf{0}$ is a N-component zero-row vector, $\Phi,\Psi,\Upsilon,\bar{\Phi},\bar{\Psi}$ and $\bar{\Upsilon}$ are $N$-component row vectors given by
\begin{eqnarray*}
\hspace{-1cm}&& \Phi=({\rm e}^{\xi_1}, {\rm e}^{\xi_2}, \cdots, {\rm e}^{\xi_N} ),\ \
\Psi=({\rm e}^{\eta_1}, {\rm e}^{\eta_2}, \cdots, {\rm e}^{\eta_N} ),\ \
\Upsilon=({\rm e}^{\chi_1}, {\rm e}^{\chi_2}, \cdots, {\rm e}^{\chi_N} ),\\
\hspace{-1cm}&& \bar{\Phi}=({\rm e}^{\bar{\xi}_1}, {\rm e}^{\bar{\xi}_2}, \cdots, {\rm e}^{\bar{\xi}_N} ),\ \
\bar{\Psi}=({\rm e}^{\bar{\eta}_1}, {\rm e}^{\bar{\eta}_2}, \cdots, {\rm e}^{\bar{\eta}_N} ),\ \
\bar{\Upsilon}=({\rm e}^{\bar{\chi}_1}, {\rm e}^{\bar{\chi}_2}, \cdots, {\rm e}^{\bar{\chi}_N} ),\ \
\end{eqnarray*}
with
\begin{eqnarray*}
\hspace{-1cm}&& \xi_i=\frac{1}{p_i-c_1}x^{(1)}_{-1}+p_ix_1+p^2_ix_2+\xi_{i0},\ \
\bar{\xi}_j=\frac{1}{\bar{p}_j+c_1}x^{(1)}_{-1}+\bar{p}_jx_1-\bar{p}^2_jx_2+\bar{\xi}_{j0},\\
\hspace{-1cm}&& \eta_i=q_i y^{(1)}_1 +\eta_{i0},\ \ \bar{\eta}_j=\bar{q}_j y^{(1)}_1 +\bar{\eta}_{j0}, \ \ \chi_i=r_i y^{(2)}_1 +\chi_{i0}, \ \ \bar{\chi}_j=\bar{r}_j y^{(2)}_1 +\bar{\chi}_{j0}.
\end{eqnarray*}
Here $p_i, \bar{p}_j, q_i, \bar{q}_j, r_i, \bar{r}_j, \xi_{i0}, \bar{\xi}_{j0}, \eta_{i0}, \bar{\eta}_{j0}, \chi_{i0}, \bar{\chi}_{j0}$ and $c_1$ are complex parameters.
Based on the Sato theory of the KP hierarchy \cite{jimbo1983solitons},
one can find that the following bilinear equations are satisfied by the above tau functions
\begin{equation}\label{mix1d-08}
\begin{array}{l}
(D_{x_2}-D^2_{x_1})\tau_{1,0}(k_1)\cdot\tau_{0,0}(k_1)=0,\\
(D_{x_2}-D^2_{x_1})\tau_{0,1}(k_1)\cdot\tau_{0,0}(k_1)=0,\\
(D_{x_2}-D^2_{x_1}-2c_1D_{x_1})\tau_{0,0}(k_1+1)\cdot\tau_{0,0}(k_1)=0,\\
D_{x_1}D_{y^{\mbox{\tiny (1)}}_1}\tau_{0,0}(k_1)\cdot\tau_{0,0}(k_1)=-2 \tau_{1,0}(k_1)\tau_{-1,0}(k_1),\\
D_{x_1}D_{y^{\mbox{\tiny (2)}}_1}\tau_{0,0}(k_1)\cdot\tau_{0,0}(k_1)=-2 \tau_{0,1}(k_1)\tau_{0,-1}(k_1),\\
(D_{x_1}D_{x^{\mbox{\tiny (1)}}_{-1}}-2)\tau_{0,0}(k_1)\cdot\tau_{0,0}(k_1)=-2 \tau_{0,0}(k_1+1)\tau_{0,0}(k_1-1).
\end{array}
\end{equation}
Here we omit the proof of the above bilinear equations, which can be given by using the Grammian technique \cite{hirota2004direct,miyake1990representation}.
Instead we carry out the reduction process to obtain bilinear form (\ref{mix1d-04}) of the (3+1)-component YO system.
By setting $x_1$, $x^{(1)}_{-1}$, $y^{(1)}_1$, $y^{(2)}_1$ are real, $x_2$, $c_1$ are pure imaginary and by letting $p^*_j=\bar{p}_j$, $q^*_j=\bar{q}_j=r^*_j=\bar{r}_j$, $\xi^*_{j0}=\bar{\xi}_{j0}$, $\eta^*_{j0}=\bar{\eta}_{j0}$ and $\chi^*_{j0}=\bar{\chi}_{j0}$, one can check that
\begin{equation*}
    a_{ij}(k_1)= a^*_{ji}(k_1),\ \  b_{ij}= b^*_{ji}.
\end{equation*}
Furthermore, we can define
\begin{eqnarray*}
&& f=\tau_{0,0}(0),\ \ g^{(1)}=\tau_{1,0}(0),\ \ g^{(2)}=\tau_{0,1}(0), \ \ h^{(1)}=\tau_{0,0}(1),\\
&& g^{(1)*}=-\tau_{-1,0}(0),\ \ g^{(2)*}=-\tau_{0,-1}(0), \ \ h^{(1)*}=\tau_{0,0}(-1),
\end{eqnarray*}
and thus the bilinear equations (\ref{mix1d-08}) become
\begin{equation}\label{mix1d-09}
\begin{array}{l}
(D_{x_2}-D^2_{x_1})g^{(k)}\cdot f=0,\\
(D_{x_2}-D^2_{x_1}-2c_1D_{x_1})h^{(1)}\cdot f=0,\\
D_{x_1}D_{y^{\tiny (k)}_1}f\cdot f=2 g^{(k)}g^{(k)*},\ \ k=1,2,\\
(D_{x_1}D_{x^{\mbox{\tiny (1)}}_{-1}}-2)f\cdot f=-2 h^{(1)}h^{(1)*}.
\end{array}
\end{equation}

Next, we conduct the dimension reduction by rewriting $f$ as
\begin{eqnarray}\label{mix1d-10}
&& f =
\left|
\begin{array}{ccc}
A' & I \\
-I & B'
\end{array}
\right|,\ \
\end{eqnarray}
via performing row and column operations, where $A'$ and $B'$ are $N\times N$ matrices defined as
\begin{eqnarray*}
\hspace{-1cm} a'_{ij}(k_1)=\frac{1}{p_i+p^*_j} ,\ \
b'_{ij}=\frac{1}{q_i+q^*_j} {\rm e}^{\eta_i+\eta^*_j +\xi^*_i+\xi_j }
+ \frac{1}{r_i+r^*_j} {\rm e}^{\chi_i+\chi^*_j+\xi^*_i+\xi_j},
\end{eqnarray*}
with
\begin{eqnarray*}
&& \eta_i+\xi^*_i=q_i y^{(1)}_1 +\frac{1}{p^*_i+c_1}x^{(1)}_{-1}+p^*_ix_1-p^{*2}_ix_2+\xi^*_{i0}+\eta_{i0},\\
&& \eta^*_j+\xi_j=q^*_j y^{(1)}_1 +\frac{1}{p_j-c_1}x^{(1)}_{-1}+p_jx_1+p^2_jx_2+\xi_{j0} +\eta^*_{j0},\\
&& \chi_i+\xi^*_i=r_i y^{(2)}_1  +\frac{1}{p^*_i+c_1}x^{(1)}_{-1}+p^*_ix_1-p^{*2}_ix_2+\xi^*_{i0}+\chi_{i0},\\
&& \chi^*_j+\xi_j=r^*_j y^{(2)}_1  +\frac{1}{p_j-c_1}x^{(1)}_{-1}+p_jx_1+p^2_jx_2+\xi_{j0} +\chi^*_{j0},
\end{eqnarray*}

Therefore, under the reduction conditions
\begin{eqnarray}
\label{mix1d-11}&& p^{*2}_i={\rm i}\sigma_1q_i-\frac{{\rm i}\sigma_3\rho^2_1}{p^*_i+c_1},\ \
p^2_j=-{\rm i}\sigma_1q^*_j+ \frac{{\rm i}\sigma_3\rho^2_1}{p_j-c_1},\\
\label{mix1d-12}&& p^{*2}_i={\rm i}\sigma_2r_i-\frac{{\rm i}\sigma_3\rho^2_1}{p^*_i+c_1},\ \
p^2_j=-{\rm i}\sigma_2r^*_j+ \frac{{\rm i}\sigma_3\rho^2_1}{p_j-c_1},
\end{eqnarray}
i.e.,
\begin{eqnarray}
\label{mix1d-13}&& \frac{1}{q_i+q^*_j}=\frac{{\rm i} \sigma_1}{ (p^*_i+p_j)[p^*_i-p_j + \frac{{\rm i}\sigma_3\rho^2_1}{(p^*_i+c_1)(p_j-c_1)} ]},\\
\label{mix1d-14}&& \frac{1}{r_i+r^*_j}=\frac{{\rm i} \sigma_2}{ (p^*_i+p_j)[p^*_i-p_j + \frac{{\rm i}\sigma_3\rho^2_1}{(p^*_i+c_1)(p_j-c_1)} ]},
\end{eqnarray}
the following relation holds
\begin{eqnarray}\label{mix1d-15}
&& \partial_{x_2} b'_{ij} = (-{\rm i} \sigma_1 \partial_{y^{(1)}_1}-{\rm i} \sigma_2 \partial_{y^{(2)}_1} +{\rm i}\sigma_3\rho^2_1 \partial_{x_{-1}}  ) b'_{ij},
\end{eqnarray}
and thus one can get
\begin{eqnarray}\label{mix1d-16}
&& f_{x_2} = -{\rm i} \sigma_1 f_{y^{(1)}_1}  -{\rm i} \sigma_2 f_{y^{(2)}_1} +{\rm i}\sigma_3\rho^2_1f_{x_{-1}},
\end{eqnarray}
and its derivative with respect to $x_1$,
\begin{eqnarray}\label{mix1d-17}
&& f_{x_1x_2} = -{\rm i} \sigma_1 f_{x_1y^{(1)}_1}  -{\rm i} \sigma_2 f_{x_1y^{(2)}_1} +{\rm i}\sigma_3\rho^2_1f_{x_1x_{-1}}.
\end{eqnarray}

On the other hand, the last three bilinear equations in (\ref{mix1d-09}) expanded as
\begin{eqnarray}\label{mix1d-18}
&& f_{x_1y^{(1)}_1 } f- f_{x_1} f_{y^{(1)}_1}=g^{(1)}g^{(1)*},\ \ f_{x_1y^{(2)}_1 } f- f_{x_1} f_{y^{(2)}_1}=g^{(2)}g^{(2)*},
\end{eqnarray}
and
\begin{eqnarray}\label{mix1d-19}
f_{x_1x^{(1)}_{-1} } f- f_{x_1} f_{x^{(1)}_{-1}}-f^2=- h^{(1)}h^{(1)*},
\end{eqnarray}
give
\begin{eqnarray}\label{mix1d-20}
\hspace{-1cm} -{\rm i}f_{x_1x_2}f+{\rm i}f_{x_2}f_{x_1}-\sigma_3\rho^2_1f^2=
-\sigma_1g^{(1)}g^{(1)*}-\sigma_2g^{(2)}g^{(2)*}-\sigma_3\rho^2_1h^{(1)}h^{(1)*},
\end{eqnarray}
by referring to Eqs. (\ref{mix1d-16})-(\ref{mix1d-17}).

Applying the variable transformations
\begin{eqnarray}\label{mix1d-21}
x_1=x,\ \ x_2=-{\rm i}t,
\end{eqnarray}
i.e.,
\begin{eqnarray*}
\partial_x=\partial_{x_1},\ \ \partial_t=-{\rm i}\partial_{x_2},
\end{eqnarray*}
and taking $c_1={\rm i}\alpha_1$, the first three equations in (\ref{mix1d-09}) become the first three bilinear equation in (\ref{mix1d-04}) and  Eq. (\ref{mix1d-20}) is nothing but the last bilinear equation in (\ref{mix1d-04}).

Lastly,
let $ {\rm e}^{\eta_{i0}}=c^{(1)*}_i$, ${\rm e}^{\eta^*_{i0}}=c^{(1)}_i$,
$ {\rm e}^{\chi_{i0}}=c^{(2)*}_i$, ${\rm e}^{\chi^*_{i0}}=c^{(2)}_i$ $(i=1,2,\cdots,N)$ and
define $ C_1=-(c^{(1)}_1,c^{(1)}_2,\cdots,c^{(1)}_N)$ and $ C_2=-(c^{(2)}_1,c^{(2)}_2,\cdots,c^{(2)}_N)$,
one can arrive at the general mixed solition (two-bright-one-dark soliton for SW components) solution to the 1D (3+1)-component YO system,
\begin{eqnarray}\label{mix1d-22}
&& f =
\left|
\begin{array}{ccc}
A & I \\
-I & B
\end{array}
\right|,\ \
%%%%%%%%%%%%
h^{(1)} =
\left|
\begin{array}{ccc}
A^{(1)} & I \\
-I & B
\end{array}
\right|,\ \
g^{(k)}=
\left|
\begin{array}{ccc}
A & I & \phi^T \\
-I & B & \textbf{0}^T\\
\textbf{0} & C_k & 0
\end{array}
\right|,
\end{eqnarray}
where $A$, $A^{(1)}$ and $B$ are $N\times N$ matrices whose entries are
\begin{eqnarray*}
&& a_{ij}=\frac{1}{p_i+p^*_j}  {\rm e}^{\theta_i+\theta^*_j},\\
&& a^{(1)}_{ij}=\frac{1}{p_i+p^*_j} \left(-\frac{p_i-{\rm i}\alpha_1}{p^*_j+{\rm i}\alpha_1} \right) {\rm e}^{\theta_i+\theta^*_j},\\
&& b_{ij}=\frac{{\rm i}\sum^2_{k=1} \sigma_k c^{(k)*}_i c^{(k)}_j }{ (p^*_i+p_j)[p^*_i-p_j + \frac{{\rm i}\sigma_3\rho^2_1}{(p^*_i+{\rm i}\alpha_1)(p_j-{\rm i}\alpha_1)} ]},
\end{eqnarray*}
and $\phi$ and $C_k$ are $N$-component row vectors given by
\begin{eqnarray*}
\hspace{-1cm}&& \phi=({\rm e}^{\theta_1}, {\rm e}^{\theta_2}, \cdots, {\rm e}^{\theta_N} ),\ \
C_k=-(c^{(k)}_1,c^{(k)}_2,\cdots,c^{(k)}_N),\ \
\end{eqnarray*}
with $\theta_i=p_ix-{\rm i}p^2_i t +\theta_{i0}$ and
 $p_i$, $\theta_{i0}$ and $c^{(k)}_i$ $(i=1,2,\cdots,N; k=1,2)$ are arbitrary complex parameters.

\subsection{One-soliton solution}

To obtain the single soliton solution, we take $N=1$ in the formula (\ref{mix1d-22}). The Gram
determinants read
\begin{eqnarray}\label{mix1d-23}
&& f =
\left|
\begin{array}{cccc}
a_{11}  & 1 \\
-1  & b_{11} \\
\end{array}
\right|,\ \
%%%%%%%%%%%%
h^{(1)} =
\left|
\begin{array}{cccc}
a^{(1)}_{11}  & 1  \\
-1  & b_{11} \\
\end{array}
\right|,
%%%%%%%%%%%%%%%
g^{(k)}=
\left|
\begin{array}{ccccc}
a_{11}  & 1  & {\rm e}^{\theta_1} \\
-1 &  b_{11}  & 0 \\
0  & -c^{(k)}_1  & 0
\end{array}
\right|,
\end{eqnarray}
where $ a_{11}=\frac{1}{p_1+p^*_1}  {\rm e}^{\theta_1+\theta^*_1},
 a^{(1)}_{11}=\frac{1}{p_1+p^*_1} \left(-\frac{p_1-{\rm i}\alpha_1}{p^*_1+{\rm i}\alpha_1} \right) {\rm e}^{\theta_1+\theta^*_1}$ and
$b_{11}=\frac{{\rm i}\sum^2_{k=1} \sigma_k c^{(k)*}_1 c^{(k)}_1 }{ (p^*_1+p_1)[p^*_1-p_1 + \frac{{\rm i}\sigma_3\rho^2_1}{(p^*_1+{\rm i}\alpha_1)(p_1-{\rm i}\alpha_1)} ]}$
for $k=1,2$.
Then one can write the above tau functions as
\begin{eqnarray}
\label{mix1d-24}&& f=1+{\rm e}^{\theta_1+\theta^*_1+2R(1,1^*) },\\
\label{mix1d-25}&& g^{(1)}=c^{(1)}_1 {\rm e}^{\theta_1 },\ \ g^{(2)}=c^{(2)}_1 {\rm e}^{\theta_1 },\\
\label{mix1d-26}&& h^{(1)}=1+{\rm e}^{\theta_1+\theta^*_1+2R(1,1^*) +2{\rm i}\phi_1},
\end{eqnarray}
with
\begin{eqnarray*}
&& {\rm e}^{2R(1,1^*)}=\frac{{\rm i}\sum^2_{k=1} \sigma_k c^{(k)*}_1 c^{(k)}_1 }{ (p^*_1+p_1)^2[p^*_1-p_1 + \frac{{\rm i}\sigma_3\rho^2_1}{(p^*_1+{\rm i}\alpha_1)(p_1-{\rm i}\alpha_1)} ]},\ \
 {\rm e}^{2{\rm i}\phi_1}=- \frac{p_1-{\rm i}\alpha_1}{p^*_1+{\rm i}\alpha_1}.
\end{eqnarray*}
It is necessary to note that this mixed soliton
solution is nonsingular only when ${\rm e}^{2R(1,1^*)}>0$.

\begin{figure}[!htbp]
\centering
{\includegraphics[height=2in,width=5.5in]{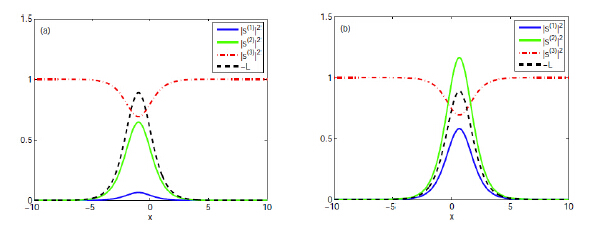}}
\caption{(Color online)  Mixed one-soliton solution (two-bright-one-dark soliton for SW components) in (3+1)-component YO system.
\label{mix1d-fig1}}
\end{figure}

The above tau functions lead to the one-soliton solution as follows
\begin{eqnarray}
\label{mix1d-27}&& S^{(k)}=\frac{c^{(k)}_1}{2} {\rm e}^{-R(1,1^*)} {\rm e}^{{\rm i}\theta_{1I}}
{\rm sech}[\theta_{1R}+R(1,1^*)],\ \ k=1,2,\\
\label{mix1d-28}&& S^{(3)}=\rho_1 {\rm e}^{{\rm i}(\alpha_1 x+ \alpha^2_1t)}
\{1+{\rm e}^{2{\rm i}\phi_1} -(1-{\rm e}^{2{\rm i}\phi_1})
{\rm tanh}[\theta_{1R}+R(1,1^*)] \},\\
\label{mix1d-29}&& L=-2p^2_{1R}{\rm sech}^2[\theta_{1R}+R(1,1^*)],
\end{eqnarray}
where $\theta_1=\theta_{1R}+{\rm i}\theta_{1I}$,
the suffixes $R$ and $I$ denote the real and imaginary parts, respectively.
The quantities $\frac{|c^{(k)}_1|}{2} {\rm e}^{-R(1,1^*)}, k=1,2$ represent the amplitude of the bright soliton in the SW
components $S^{(k)}$ and the real quantity $2p^2_{1R}$ denotes the amplitude of soliton in the LW component $-L$.
For the dark soliton in the SW component $S^{(3)}$, $|S^{(3)}|$ approaches $|\rho_1|$ as $x\rightarrow \pm \infty$,
and the intensity is $|\rho_1|\cos\phi_1$.
As an example, we illustrate the mixed one-soliton at time $t=0$ in Fig. \ref{mix1d-fig1} for the nonlinearity coefficients $(\sigma_1,\sigma_2,\sigma_3)=(1,-1,1)$.
The parameters are chosen as $\rho_1=1$, $\alpha_1=2$, $p_1=\frac{2}{3}+{\rm i}$, $\theta_{10}=0$, $c^{(1)}_1=1$ and (a) $c^{(2)}_1=3+{\rm i}$; (b) $c^{(2)}_1=1+{\rm i}$.
One can observe that when the parameters $c^{(k)}_1$ take different values,
the intensities of bright solitons for SW components $S^{(1)}$ and $S^{(2)}$ change, but the depths of dark soliton for SW component $S^{(3)}$ and of the soliton for LW component $L$ remain the same.

\subsection{Two-soliton solution}

By taking $N=2$ in (\ref{mix1d-22}), we have the following Gram determinants
\begin{eqnarray}
\label{mix1d-30}&& f =
\left|
\begin{array}{cccc}
a_{11} & a_{12} & 1 & 0 \\
a_{21} & a_{22} & 0 & 1 \\
-1 & 0 & b_{11} & b_{12} \\
0 & -1 & b_{21} & b_{22} \\
\end{array}
\right|,\ \
%%%%%%%%%%%%
h^{(1)} =
\left|
\begin{array}{cccc}
a^{(1)}_{11} & a^{(1)}_{12} & 1 & 0 \\
a^{(1)}_{21} & a^{(1)}_{22} & 0 & 1 \\
-1 & 0 & b_{11} & b_{12} \\
0 & -1 & b_{21} & b_{22} \\
\end{array}
\right|,\\
%%%%%%%%%%%%%%%
\label{mix1d-31}&&g^{(k)}=
\left|
\begin{array}{ccccc}
a_{11} & a_{12} & 1 & 0 & {\rm e}^{\theta_1} \\
a_{21} & a_{22} & 0 & 1 & {\rm e}^{\theta_2}\\
-1 & 0 & b_{11} & b_{12} & 0 \\
0 & -1 & b_{21} & b_{22} &0 \\
0 & 0 & -c^{(k)}_1 & -c^{(k)}_2 & 0
\end{array}
\right|,
\end{eqnarray}
where $a_{ij}=\frac{1}{p_i+p^*_j}  {\rm e}^{\theta_i+\theta^*_j}$,
$ a^{(1)}_{ij}=\frac{1}{p_i+p^*_j} \left(-\frac{p_i-{\rm i}\alpha_1}{p^*_j+{\rm i}\alpha_1} \right) {\rm e}^{\theta_i+\theta^*_j}$ and
$ b_{ij}=\frac{{\rm i}\sum^2_{k=1} \sigma_k c^{(k)*}_i c^{(k)}_j }{ (p^*_i+p_j)[p^*_i-p_j + \frac{{\rm i}\sigma_3\rho^2_1}{(p^*_i+{\rm i}\alpha_1)(p_j-{\rm i}\alpha_1)} ]}$
for $k=1,2$.
Then the explicit form of the above tau functions can be expressed as
\begin{eqnarray}
\label{mix1d-32}\hspace{-2cm}\nonumber  f&=&1+E(1,1^*){\rm e}^{\theta_1+\theta^*_1} + E(1,2^*){\rm e}^{\theta_1+\theta^*_2}
+ E(2,1^*){\rm e}^{\theta_2+\theta^*_1} +E(2,2^*){\rm e}^{\theta_2+\theta^*_2}\\
\label{mix1d-33}\hspace{-2cm}  &&+E(1,1^*,2,2^*){\rm e}^{\theta_1+\theta_2+\theta^*_1+\theta^*_2},\\
\hspace{-2cm}  g^{(k)}&=&c^{(k)}_1{\rm e}^{\theta_1} + c^{(k)}_2{\rm e}^{\theta_2}
+F^{(k)}(1,2,1^*){\rm e}^{\theta_1+\theta_2+\theta^*_1}+F^{(k)}(1,2,2^*){\rm e}^{\theta_1+\theta_2+\theta^*_2},\\
\label{mix1d-34}\hspace{-2cm}\nonumber  h^{(1)}&=&1+G(1,1^*){\rm e}^{\theta_1+\theta^*_1} + G(1,2^*){\rm e}^{\theta_1+\theta^*_2}
+ G(2,1^*){\rm e}^{\theta_2+\theta^*_1} +G(2,2^*){\rm e}^{\theta_2+\theta^*_2}\\
\hspace{-2cm}  &&+G(1,1^*,2,2^*){\rm e}^{\theta_1+\theta_2+\theta^*_1+\theta^*_2},
\end{eqnarray}
where
\begin{eqnarray*}
\hspace{-2cm} && E(i,j^*)=\frac{{\rm i}\sum^2_{k=1} \sigma_k c^{(k)*}_i c^{(k)}_j }{ (p_i+p^*_j)^2[p^*_i-p_j + \frac{{\rm i}\sigma_3\rho^2_1}{(p^*_i+{\rm i}\alpha_1)(p_j-{\rm i}\alpha_1)} ]},\ \
G(i,j^*)=-\frac{p_i-{\rm i}\alpha_1}{p^*_j+{\rm i}\alpha_1} E(i,j^*),\\
\hspace{-2cm} && E(1,1^*,2,2^*)=|p_1-p_2|^2\left[\frac{E(1,1^*)E(2,2^*)}{(p_1+p^*_2)(p_2+p^*_1)}
- \frac{E(1,2^*)E(2,1^*)}{(p_1+p^*_1)(p_2+p^*_2)} \right],\\
\hspace{-2cm} &&G(1,1^*,2,2^*)=\frac{(p_1-{\rm i}\alpha_1)(p_2-{\rm i}\alpha_1)}{(p^*_1+{\rm i}\alpha_1)(p^*_2+{\rm i}\alpha_1)}E(1,1^*,2,2^*),\\
\hspace{-2cm} && F^{(k)}(1,2,i^*)=(p_2-p_1) \left[ \frac{c^{(k)}_2E(1,i^*)}{p_2+p^*_i}-  \frac{c^{(k)}_1E(2,i^*)}{p_1+p^*_i} \right],
\end{eqnarray*}
and $\theta_i=p_ix-{\rm i}p^2_i t +\theta_{i0}$ for $i=1,2$.
%%%%%%%%%%%%%%%%%%%%%%%%%%%%%%%%%%

\begin{figure}[!htbp]
\centering
{\includegraphics[height=1.3in,width=6in]{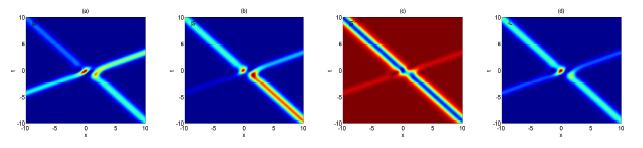}}
\caption{(Color online) Mixed two-soliton solution (two-bright-one-dark soliton for SW components) including inelastic collisions for the SW components $S^{(1)}$ and $S^{(2)}$ in the (3+1)-component YO system
.\label{mix1d-fig2}}
\end{figure}

\begin{figure}[!htbp]
\centering
{\includegraphics[height=1.3in,width=6in]{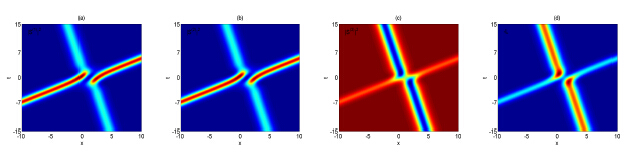}}
\caption{(Color online) Mixed two-soliton solution (two-bright-one-dark soliton for SW components) including elastic collisions for the SW components $S^{(1)}$ and $S^{(2)}$ in the (3+1)-component YO system
.\label{mix1d-fig3}}
\end{figure}

The above solution contains both singular and nonsingular solutions.
To assure the nonsingular solution, the denominator $f$ needs to be real and nonzero.
The expression for $f$ can be rewritten as
\begin{eqnarray}
\label{mix1d-35}\hspace{-2cm}\nonumber f&=&2{\rm e}^{\theta_{1R}+\theta_{2R}}
[{\rm e}^{\frac{\Omega_1+\Omega_2}{2}} {\rm cosh}(\theta_{1R}-\theta_{2R}+\frac{\Omega_1-\Omega_2}{2})
+{\rm e}^{H_{1R}} \cos(\theta_{1I}-\theta_{2I}+H_{1I})\\
\hspace{-2cm} &&+{\rm e}^{\frac{\Omega_3}{2}} {\rm cosh}(\theta_{1R}-\theta_{2R}+\frac{\Omega_3}{2})],
\end{eqnarray}
where
\begin{eqnarray*}
\hspace{-2cm}&& {\rm e}^{\Omega_1}=E(1,1^*),\ \ {\rm e}^{\Omega_2}=E(2,2^*),\ \ {\rm e}^{\Omega_3}=E(1,1^*,2,2^*),\ \
{\rm e}^{H_{1R}+{\rm i}H_{1I}}=E(1,2^*).
\end{eqnarray*}
Combining the requirement for the existence of one-soliton solution,
we can conclude that the condition $E(i ,i^*)>0,i=1,2$ is a necessary condition
and ${\rm e}^{\frac{\Omega_1+\Omega_2}{2}}+{\rm e}^{\frac{\Omega_3}{2}}>{\rm e}^{H_{1R}}$ is a sufficient condition to guarantee a
nonsingular two-soliton solution.
For the interaction properties of these solitons, one can carry out the asymptotic analysis as in \cite{kanna2013general,kanna2009higher,kanna2014mixed},
and deduce that the bright solitons appearing in SW components $S^{(i)}(i=1,2)$
only undertake elastic collisions under some special parameters,
while the dark solitons in the SW component $S^{(3)}$ and the bright soliton in the LW component $-L$ always exhibit elastic collisions.
More precisely, bright solitons in SW components $S^{(i)}(i=1,2)$ undergo elastic collisions
if the parameters satisfy the condition $\frac{c^{(1)}_1}{c^{(1)}_2}=\frac{c^{(2)}_1}{c^{(2)}_2}$.
Otherwise, they undertake inelastic collisions (shape changing).
For illustrative purpose, the interactions of two solitons are displayed in
Figs. \ref{mix1d-fig2} and \ref{mix1d-fig3} for the (3+1)-component YO system with the nonlinearity coefficients $(\sigma_1,\sigma_2,\sigma_3)=(1,-1,1)$.
The parameters in Fig.\ref{mix1d-fig2} are chosen as $\rho_1=1, \alpha_1=1,
p_1=\frac{2}{3}-\frac{5}{4}{\rm i}, p_2=1+\frac{1}{2}{\rm i},
c^{(1)}_1=1+{\rm i},c^{(1)}_2=2, c^{(2)}_1=\frac{1}{2}{\rm i},c^{(2)}_2=1+2{\rm i}$
and $\theta_{10}=\theta_{20}=0$, which result in inelastic collisions for the bright solitons in the SW components $S^{(i)}(i=1,2)$.
Meanwhile, we show an example of elastic collision between bright solitons in two SW components in Fig.\ref{mix1d-fig3}
with the parameters $\rho_1=1, \alpha_1=1,
p_1=\frac{2}{3}-\frac{3}{4}{\rm i}, p_2=1+\frac{1}{10}{\rm i},
c^{(1)}_1=2,c^{(1)}_2=1, c^{(2)}_1=\frac{3}{2},c^{(2)}_2=\frac{3}{4}$
and $\theta_{10}=\theta_{20}=0$.
In Figs. \ref{mix1d-fig2} and \ref{mix1d-fig3}, the SW components $S^{(1)}$ and $S^{(2)}$ display two bright solitons collisions in (a) and (b), the SW component $S^{(3)}$ shows two dark solitons collision in (c), and (d) represents two bright solitons collision of the LW component $-L$.
As shown in (a) and (b) for the SW components collision of $S^{(1)}$ and $S^{(2)}$, respectively, one is elastic collision (\ref{mix1d-fig2}), the other is inelastic collision (\ref{mix1d-fig3}).

%%%%%%%%%%%%%%%%%%%%%%%%%%%%%%%%%%
\begin{figure}[!htbp]
\centering
{\includegraphics[height=1.3in,width=6in]{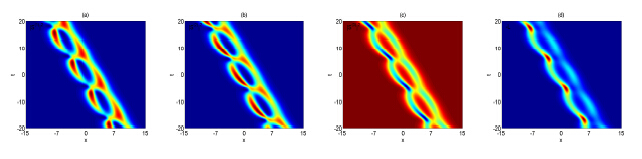}}
\caption{(Color online) Mixed two-soliton bound state (two-bright-one-dark soliton for SW components) in the (3+1)-component YO system.
\label{mix1d-fig4}}
\end{figure}

In addition, soliton bound states are one class of special multisoliton solutions,
in which multiple solitons move with the same velocity.
By assuming $p_i=p_{iR}+{\rm i}p_{iI}$, one can obtain the mixed two-soliton bound state
from (\ref{mix1d-32})-(\ref{mix1d-34}) with the restriction $p_{1I}=p_{2I}$.
To demonstrate such a bound state, we choose the parameters as
$\rho_1=1, \alpha_1=1,
p_1=\frac{2}{3}+\frac{1}{5}{\rm i}, p_2=1+\frac{1}{5}{\rm i},
c^{(1)}_1=2+{\rm i},c^{(1)}_2=2+4{\rm i}, c^{(2)}_1=\frac{1}{3}+{\rm i},c^{(2)}_2=\frac{1}{2}+2{\rm i},
\theta_{10}=\theta_{20}=0$ under the same nonlinearity coefficients as in Figs. \ref{mix1d-fig2}
and show the contour plots of all components in Fig. \ref{mix1d-fig4}.
Here, Fig. \ref{mix1d-fig4} (a) and (b) exhibit two-bright-soliton bound states for the SW components $S^{(1)}$ and $S^{(2)}$ with the different amplitudes,
Fig. \ref{mix1d-fig4} (c) displays two-dark-soliton bound state for the SW component $S^{(3)}$ and two-bright-soliton bound state of the LW component $-L$ is shown in  Fig. \ref{mix1d-fig4} (d).

\subsection{One-bright-two-dark soliton for the SW components}

In this subsection, we assume the SW component $S^{(1)}$ is of bright type and the SW components $S^{(2)}$ and $S^{(3)}$ are of dark type.
The dependent variable transformations
\begin{equation}\label{mix1d-36}
\begin{array}{l}
\hspace{-2cm} S^{(1)}=\frac{g^{(1)}}{f},\ \ S^{(2)}=\rho_1\frac{h^{(1)}}{f} {\rm e}^{{\rm i}(\alpha_1 x+\alpha^2_1 t)},\ \  S^{(3)}=\rho_2\frac{h^{(2)}}{f} {\rm e}^{{\rm i}(\alpha_2 x +\alpha^2_2 t)} ,\ \ L=-2(\log f)_{xx},\ \
\end{array}
\end{equation}
convert the (3+1)-component YO Eqs.(\ref{mix1d-02}) into the following bilinear equations:
\begin{equation}\label{mix1d-37}
\begin{array}{l}
\left[ \textmd{i} D_t-D^2_x \right]g^{(1)}\cdot f=0,\\
\left[ \textmd{i} (D_t-2\alpha_lD_x)-D^2_x \right]h^{(l)}\cdot f=0,\ \ l=1,2, \\
\left(D_tD_x-2 \sum^2_{l=1}\sigma_{l+1}\rho^2_l \right)f\cdot f= -2\sigma_1 g^{(1)}g^{(1)*}-2\sum^2_{l=1}\sigma_{l+1}\rho^2_l h^{(l)}h^{(l)*},
\end{array}
\end{equation}
where $g^{(1)},h^{(1)}$ and $h^{(2)}$ are complex-valued functions, $f$ is a real-valued function,
$\alpha_l$ and $\rho_l (\rho_l>0; l=1,2)$ are arbitrary real constants.

Here the bilinear form of the (3+1)-component YO system (\ref{mix1d-37}) is viewed as a reduction of two-component KP hierarchy.
To this end, we start with the tau functions expressed in Gram determinants as follows
\begin{eqnarray}
\label{mix1d-38}&& \tau_{0}(k_1,k_2) =
\left|
\begin{array}{ccc}
A & I \\
-I & B
\end{array}
\right|,\\
\label{mix1d-39}&& \tau_{1}(k_1,k_2)=
\left|
\begin{array}{ccc}
A & I & \Phi^T \\
-I & B & \textbf{0}^T\\
\textbf{0} & -\bar{\Psi} & 0
\end{array}
\right|,\ \
 \tau_{-1}(k_1,k_2)=
\left|
\begin{array}{ccc}
A & I & \textbf{0}^T \\
-I & B & \Psi^T\\
-\bar{\Phi} & \textbf{0} & 0
\end{array}
\right|,
\end{eqnarray}
where $\Phi,\Psi,\bar{\Phi},\bar{\Psi}$  are $N$-component row vectors defined previously, $A$ and $B$ are $N\times N$ matrices whose entries are
\begin{eqnarray*}
\hspace{-1.5cm} a_{ij}(k_1,k_2)=\frac{1}{p_i+\bar{p}_j} \left(-\frac{p_i-c_1}{\bar{p}_j+c_1} \right)^{k_1}  \left(-\frac{p_i-c_2}{\bar{p}_j+c_2} \right)^{k_2} {\rm e}^{\xi_i+\bar{\xi}_j},\ \
b_{ij}=\frac{1}{q_i+\bar{q}_j} {\rm e}^{\eta_i+\bar{\eta}_j} ,
\end{eqnarray*}
with
\begin{eqnarray*}
\hspace{-1cm}&& \xi_i=\frac{1}{p_i-c_1}x^{(1)}_{-1}+\frac{1}{p_i-c_2}x^{(2)}_{-1}+p_ix_1+p^2_ix_2+\xi_{i0},\\
&&\bar{\xi}_j=\frac{1}{\bar{p}_j+c_1}x^{(1)}_{-1}+\frac{1}{\bar{p}_j+c_2}x^{(2)}_{-1}+\bar{p}_jx_1-\bar{p}^2_jx_2+\bar{\xi}_{j0},\\
\hspace{-1cm}&& \eta_i=q_i y^{(1)}_1 +\eta_{i0},\ \ \bar{\eta}_j=\bar{q}_j y^{(1)}_1 +\bar{\eta}_{j0}.
\end{eqnarray*}
Here $p_i, \bar{p}_j, q_i, \bar{q}_j, \xi_{i0}, \bar{\xi}_{j0}, \eta_{i0}, \bar{\eta}_{j0}, c_1$ and $c_2$ are complex parameters.
Based on the Sato theory for the KP hierarchy \cite{jimbo1983solitons},
the above tau functions satisfy the bilinear equations
\begin{equation}\label{mix1d-40}
\begin{array}{l}
(D_{x_2}-D^2_{x_1})\tau_{1}(k_1,k_2)\cdot\tau_{0}(k_1,k_2)=0,\\
(D_{x_2}-D^2_{x_1}-2c_1D_{x_1})\tau_{0}(k_1+1,k_2)\cdot\tau_{0}(k_1,k_2)=0,\\
(D_{x_2}-D^2_{x_1}-2c_2D_{x_1})\tau_{0}(k_1,k_2+1)\cdot\tau_{0}(k_1,k_2)=0,\\
D_{x_1}D_{y^{(1)}_1}\tau_{0}(k_1,k_2)\cdot\tau_{0}(k_1,k_2)=-2 \tau_{1}(k_1,k_2)\tau_{-1}(k_1,k_2),\\
(D_{x_1}D_{x^{\mbox{\tiny (1)}}_{-1}}-2)\tau_{0}(k_1,k_2)\cdot\tau_{0}(k_1,k_2)=-2 \tau_{0}(k_1+1,k_2)\tau_{0}(k_1-1,k_2),\\
(D_{x_1}D_{x^{\mbox{\tiny (2)}}_{-1}}-2)\tau_{0}(k_1,k_2)\cdot\tau_{0}(k_1,k_2)=-2 \tau_{0}(k_1,k_2+1)\tau_{0}(k_1,k_2-1),
\end{array}
\end{equation}
Next we perform the reduction process to obtain the bilinear equations (\ref{mix1d-37}).
We first consider the complex conjugate reduction
by setting $x_1$, $x^{(1)}_{-1}$, $x^{(2)}_{-1}$, $y^{(1)}_1$ to be real, $x_2$, $c_1$ $c_2$ to be pure imaginary and by letting $p^*_j=\bar{p}_j$, $q^*_j=\bar{q}_j$, $\xi^*_{j0}=\bar{\xi}_{j0}$, and $\eta^*_{j0}=\bar{\eta}_{j0}$.
Then, it is easy to see that
\begin{equation*}
 a_{ij}(k_1,k_2)=a^*_{ji}(-k_1,-k_2),\ \ b_{ij}=b^*_{ji}.
\end{equation*}
Thus, it then follows
\begin{eqnarray*}
&& f=\tau_{0}(0,0),\ \ g^{(1)}=\tau_{1}(0,0), \ \ h^{(1)}=\tau_{0}(1,0),\ \ h^{(2)}=\tau_{0}(0,1), \\
&& g^{(1)*}=-\tau_{-1}(0,0), \ \ h^{(1)*}=\tau_{0}(-1,0),\ \ h^{(2)*}=\tau_{0}(0,-1),
\end{eqnarray*}
the bilinear equations (\ref{mix1d-40}) become
\begin{equation}\label{mix1d-41}
\begin{array}{l}
(D_{x_2}-D^2_{x_1})g^{(1)}\cdot f=0,\\
(D_{x_2}-D^2_{x_1}-2c_lD_{x_1})h^{(l)}\cdot f=0,\\
D_{x_1}D_{y^{(1)}_1}f\cdot f=2 g^{(1)}g^{(1)*},\\
(D_{x_1}D_{x^{\mbox{\tiny (l)}}_{-1}}-2)f\cdot f=-2 h^{(l)}h^{(l)*},\ \ l=1,2.
\end{array}
\end{equation}

Similar to the two-bright-one-dark soliton case, we can show that if $q_i$ satisfies
\begin{eqnarray}
\label{mix1d-42}&& p^{*2}_i={\rm i}\sigma_1q_i-\frac{{\rm i}\sigma_2\rho^2_1}{p^*_i+c_1} -\frac{{\rm i}\sigma_3\rho^2_2}{p^*_i+c_2},\\
\label{mix1d-43}&& p^2_j=-{\rm i}\sigma_1q^*_j + \frac{{\rm i}\sigma_2\rho^2_1}{p_j-c_1} + \frac{{\rm i}\sigma_3\rho^2_2}{p_j-c_2},
\end{eqnarray}
i.e.,
\begin{eqnarray}
\label{mix1d-44}&& \frac{1}{q_i+q^*_j}=\frac{{\rm i} \sigma_1}{ (p^*_i+p_j)
[p^*_i-p_j + \frac{{\rm i}\sigma_2\rho^2_1}{(p^*_i+c_1)(p_j-c_1)}  + \frac{{\rm i}\sigma_3\rho^2_2}{(p^*_i+c_2)(p_j-c_2)} ]},
\end{eqnarray}
one can get the following relation
\begin{eqnarray}
\label{mix1d-45}&& f_{x_2} = -{\rm i} \sigma_1 f_{y^{(1)}_1}  +{\rm i}\sigma_2\rho^2_1f_{x^{(1)}_{-1}} +{\rm i}\sigma_3\rho^2_2f_{x^{(2)}_{-1}} ,
\end{eqnarray}
and its derivative with respect to $x_1$,
\begin{eqnarray}
\label{mix1d-46}&& f_{x_1x_2} = -{\rm i} \sigma_1 f_{x_1y^{(1)}_1}
+{\rm i}\sigma_2\rho^2_1f_{x_1x^{(1)}_{-1}} +{\rm i}\sigma_3\rho^2_2f_{x_1x^{(2)}_{-1}}.
\end{eqnarray}

On the other hand, the last three bilinear equations in (\ref{mix1d-41}) expanded as
\begin{eqnarray}
\label{mix1d-47}&& f_{x_1y^{(1)}_1 } f- f_{x_1} f_{y^{(1)}_1}=g^{(1)}g^{(1)*},\ \
\end{eqnarray}
and
\begin{eqnarray}
\label{mix1d-48}\hspace{-1cm} f_{x_1x^{(1)}_{-1} } f- f_{x_1} f_{x^{(1)}_{-1}}-f^2=- h^{(1)}h^{(1)*},\ \
f_{x_1x^{(2)}_{-1} } f- f_{x_1} f_{x^{(2)}_{-1}}-f^2=- h^{(2)}h^{(2)*},
\end{eqnarray}
give
\begin{eqnarray}
\label{mix1d-49}\hspace{-2.3cm} -{\rm i}f_{x_1x_2}f+{\rm i}f_{x_2}f_{x_1}-(\sigma_2\rho^2_1+\sigma_3\rho^2_2)f^2=
-\sigma_1g^{(1)}g^{(1)*}-\sigma_2\rho^2_1h^{(1)}h^{(1)*} -\sigma_3\rho^2_2h^{(2)}h^{(2)*},
\end{eqnarray}
by referring to Eqs. (\ref{mix1d-45})-(\ref{mix1d-46}). Eq.(\ref{mix1d-49}) is exactly the last bilinear equation in (\ref{mix1d-37}) through the variable transformation (\ref{mix1d-21}) and $c_1={\rm i}\alpha_1,c_2={\rm i}\alpha_2$. Under the same transformation, the first three bilinear equations in (\ref{mix1d-41})
become the first three in (\ref{mix1d-37}).

In summary, we complete the reductions from the bilinear
equations in (\ref{mix1d-40}) to the ones in (\ref{mix1d-37}). Thus, we are able to construct general mixed soliton (one-bright-two-dark soliton for SW components) solution to the the 1D (3+1)-component YO system,
\begin{eqnarray}\label{mix1d-50}
&& f =
\left|
\begin{array}{ccc}
A & I \\
-I & B
\end{array}
\right|,\ \
%%%%%%%%%%%%
h^{(l)} =
\left|
\begin{array}{ccc}
A^{(l)} & I \\
-I & B
\end{array}
\right|,\ \
g^{(1)}=
\left|
\begin{array}{ccc}
A & I & \phi^T \\
-I & B & \textbf{0}^T\\
\textbf{0} & C_1 & 0
\end{array}
\right|,
\end{eqnarray}
where $A$, $A^{(l)}$ and $B$ are $N\times N$ matrices whose entries are
\begin{eqnarray*}
&& a_{ij}=\frac{1}{p_i+p^*_j}  {\rm e}^{\theta_i+\theta^*_j},\\
&& a^{(l)}_{ij}=\frac{1}{p_i+p^*_j} \left(-\frac{p_i-{\rm i}\alpha_l}{p^*_j+{\rm i}\alpha_l} \right) {\rm e}^{\theta_i+\theta^*_j},\\
&& b_{ij}=\frac{{\rm i} \sigma_1 c^{(1)*}_i c^{(1)}_j }{ (p^*_i+p_j)
[p^*_i-p_j + \frac{{\rm i}\sigma_2\rho^2_1}{(p^*_i+{\rm i}\alpha_1)(p_j-{\rm i}\alpha_1)}  + \frac{{\rm i}\sigma_3\rho^2_2}{(p^*_i+{\rm i}\alpha_2)(p_j-{\rm i}\alpha_2)} ]},
\end{eqnarray*}
and $\phi$ and $C_1$ are $N$-component row vectors given by
\begin{eqnarray*}
\hspace{-1cm}&& \phi=({\rm e}^{\theta_1}, {\rm e}^{\theta_2}, \cdots, {\rm e}^{\theta_N} ),\ \
C_1=-(c^{(1)}_1,c^{(1)}_2,\cdots,c^{(1)}_N),\ \
\end{eqnarray*}
with $\theta_i=p_ix-{\rm i}p^2_i t +\theta_{i0}$ and
 $p_i$, $\theta_{i0}$ and $c^{(1)}_i={\rm e}^{\eta_{i0}}$ $(i=1,2,\cdots,N; l=1,2)$ are arbitrary complex parameters.

\textbf{Remark 3.1}
In the last subsection, we have constructed the soliton solution in which the SW components $S^{(1)}$ and $S^{(2)}$ are of bright type and the SW component $S^{(3)}$ is of dark type in the (3+1)-component YO system. It is noted that we start from the (2+1)-component KP hierarchy with one copy of shifted singular point ($c$).
In contrast, for the soliton solution (the SW components $S^{(1)}$ is of bright type and the SW component $S^{(2)}$ and $S^{(3)}$ are of dark type) obtained in this subsection, we begin with the (1+1)-component KP hierarchy with two copies of shifted singular points ($c_1$ and $c_2$).
Thus the number of the components in KP hierarchy matches the numbers of the components possessing the bright soliton
while the number of the copies of shifted singular points coincides with the numbers of the components possessing the dark soliton.
This fact can be refered to the construction in Ref.\cite{jimbo1983solitons}.

%%%%%%%%%%%%%%%%%%%%%%%%%%%%%%%%%%
\subsection{One-soliton solution}
By taking $N=1$ in the formula (\ref{mix1d-50}), we get the Gram
determinants
\begin{eqnarray}\label{mix1d-51}
&& f =
\left|
\begin{array}{cccc}
a_{11}  & 1 \\
-1  & b_{11} \\
\end{array}
\right|,\ \
%%%%%%%%%%%%
h^{(l)} =
\left|
\begin{array}{cccc}
a^{(l)}_{11}  & 1  \\
-1  & b_{11} \\
\end{array}
\right|,\ \
%%%%%%%%%%%%%%%
g^{(1)}=
\left|
\begin{array}{ccccc}
a_{11}  & 1  & {\rm e}^{\theta_1} \\
-1 &  b_{11}  & 0 \\
0  & -c^{(1)}_1  & 0
\end{array}
\right|,
\end{eqnarray}
where $a_{11}=\frac{1}{p_1+p^*_1}  {\rm e}^{\theta_1+\theta^*_1},
 a^{(l)}_{11}=\frac{1}{p_1+p^*_1} \left(-\frac{p_1-{\rm i}\alpha_l}{p^*_1+{\rm i}\alpha_l} \right) {\rm e}^{\theta_1+\theta^*_1}$
 and
$b_{11}=\frac{{\rm i} \sigma_1 c^{(1)*}_1 c^{(1)}_1 }{ (p^*_1+p_1)\Delta_{11}
}$, $\Delta_{11}=[p^*_1-p_1 + \sum^2_{l=1}\frac{{\rm i}\sigma_{l+1}\rho^2_l}{(p^*_1+{\rm i}\alpha_l)(p_1-{\rm i}\alpha_l)}  ]$
for $l=1,2$.
These tau functions can be rewritten as
\begin{eqnarray}
\label{mix1d-52} && f=1+{\rm e}^{\theta_1+\theta^*_1+2R(1,1^*) },\ \
 g^{(1)}=c^{(1)}_1 {\rm e}^{\theta_1 },\\
\label{mix1d-53} && h^{(l)}=1+{\rm e}^{\theta_1+\theta^*_1+2R(1,1^*) +2{\rm i}\phi_l},\ \ l=1,2,
\end{eqnarray}
with
\begin{eqnarray*}
&& {\rm e}^{2R(1,1^*)}=\frac{{\rm i} \sigma_1 c^{(1)*}_1 c^{(1)}_1 }{ (p^*_1+p_1)^2\Delta_{11}
},\ \
 {\rm e}^{2{\rm i}\phi_l}=- \frac{p_1-{\rm i}\alpha_l}{p^*_1+{\rm i}\alpha_l},
\end{eqnarray*}
where $\Delta_{11}=[p^*_1-p_1 + \sum^2_{l=1}\frac{{\rm i}\sigma_{l+1}\rho^2_l}{(p^*_1+{\rm i}\alpha_l)(p_1-{\rm i}\alpha_l)}  ]$.
Note that this mixed soliton
solution is nonsingular only when ${\rm e}^{2R(1,1^*)}>0$.

\begin{figure}[!htbp]
\centering
{\includegraphics[height=2in,width=5.5in]{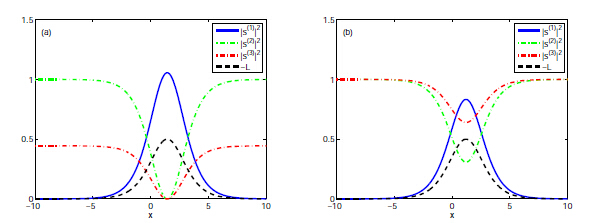}}
\caption{(Color online) Mixed one-soliton solution (one-bright-two-dark soliton for SW components) in (3+1)-component YO system.
\label{mix1d-fig5}}
\end{figure}

In this case, the one-soliton solution has the following form
\begin{eqnarray}
\label{mix1d-54}\hspace{-1cm}&& S^{(1)}=\frac{c^{(1)}_1}{2} {\rm e}^{-R(1,1^*)} {\rm e}^{{\rm i}\theta_{1I}}
{\rm sech}[\theta_{1R}+R(1,1^*)],\\
\label{mix1d-55}\hspace{-1cm}&& S^{(l+1)}=\rho_l {\rm e}^{{\rm i}(\alpha_l x+ \alpha^2_lt)}
\{1+{\rm e}^{2{\rm i}\phi_l} -(1-{\rm e}^{2{\rm i}\phi_l})
{\rm tanh}[\theta_{1R}+R(1,1^*)] \},\ \ l=1,2,\ \ \\
\label{mix1d-56}\hspace{-1cm}&& L=-2p^2_{1R}{\rm sech}^2[\theta_{1R}+R(1,1^*)],
\end{eqnarray}
where $\theta_1=\theta_{1R}+{\rm i}\theta_{1I}$,
the suffixes $R$ and $I$ denote the real and imaginary parts, respectively.
The quantity $\frac{|c^{(1)}_1|}{2} {\rm e}^{-R(1,1^*)}$ represents the amplitude of the bright soliton in the SW
components $S^{(1)}$ and the real quantity $2p^2_{1R}$ denotes the amplitude of the soliton in the LW component $-L$.
For the dark soliton in the SW components $S^{(2)}$ and $S^{(3)}$, $|S^{(l+1)}|$ approaches $|\rho_l|$ as $x\rightarrow \pm \infty$,
and the intensity is $|\rho_l|\cos\phi_l$ for $l=1,2$.
When $\alpha_1$ and $\alpha_2$ take different values, there are two cases:
(i) $\alpha_1=\alpha_2$.
In this case, $\phi_1=\phi_2$ means dark solitons in SW components $S^{(2)}$ and $S^{(3)}$ are
proportional to each other.
Thus this situation is viewed as degenerate case for dark solitons.
(ii) $\alpha_1\neq\alpha_2$.
The condition $\phi_1\neq\phi_2$ implies that dark solitons in SW components $S^{(2)}$ and $S^{(3)}$
have different degrees of darkness at the center.
In this situation, the SW components $S^{(2)}$ and $S^{(3)}$ are not proportional to
each other.
We illustrate such  degenerate and non-degenerate cases for the choice of the nonlinearity coefficients $(\sigma_1,\sigma_2,\sigma_3)=(1,1,-1)$ in Fig. \ref{mix1d-fig5}.
The parameters are chosen as $\rho_1=1$, $p_1=\frac{1}{2}-{\rm i}$, $\theta_{10}=0$, $c^{(1)}_1=1$ and (a) $\alpha_1=\alpha_2=-1$, $\rho_2=\frac{1}{2}$; (b) $\alpha_1=-\frac{2}{3},\alpha_2=-\frac{1}{3}$, $\rho_2=1$  at time $t=0$.

%%%%%%%%%%%%%%%%%%%%%%%%%%%%%%%%%%%%%%%%%%%%%%%%
\subsection{Two-soliton solution}
When $N=2$ in (\ref{mix1d-50}),  the Gram determinants are of the form
\begin{eqnarray}
\label{mix1d-57}&& f =
\left|
\begin{array}{cccc}
a_{11} & a_{12} & 1 & 0 \\
a_{21} & a_{22} & 0 & 1 \\
-1 & 0 & b_{11} & b_{12} \\
0 & -1 & b_{21} & b_{22} \\
\end{array}
\right|,\ \
%%%%%%%%%%%%
h^{(l)} =
\left|
\begin{array}{cccc}
a^{(l)}_{11} & a^{(l)}_{12} & 1 & 0 \\
a^{(l)}_{21} & a^{(l)}_{22} & 0 & 1 \\
-1 & 0 & b_{11} & b_{12} \\
0 & -1 & b_{21} & b_{22} \\
\end{array}
\right|,\\
%%%%%%%%%%%%%%%
\label{mix1d-58}&&g^{(1)}=
\left|
\begin{array}{ccccc}
a_{11} & a_{12} & 1 & 0 & {\rm e}^{\theta_1} \\
a_{21} & a_{22} & 0 & 1 & {\rm e}^{\theta_2}\\
-1 & 0 & b_{11} & b_{12} & 0 \\
0 & -1 & b_{21} & b_{22} &0 \\
0 & 0 & -c^{(1)}_1 & -c^{(1)}_2 & 0
\end{array}
\right|,
\end{eqnarray}
where $a_{ij}=\frac{1}{p_i+p^*_j}  {\rm e}^{\theta_i+\theta^*_j},
 a^{(l)}_{ij}=\frac{1}{p_i+p^*_j} \left(-\frac{p_i-{\rm i}\alpha_l}{p^*_j+{\rm i}\alpha_l} \right) {\rm e}^{\theta_i+\theta^*_j}$
 and
$b_{ij}=\frac{{\rm i} \sigma_1 c^{(1)*}_i c^{(1)}_j }{ (p^*_i+p_j)\Delta_{ij}
}$, $\Delta_{ij}=[p^*_i-p_j + \sum^2_{l=1}\frac{{\rm i}\sigma_{l+1}\rho^2_l}{(p^*_i+{\rm i}\alpha_l)(p_j-{\rm i}\alpha_l)}  ]$
for $l=1,2$.
Then we can express the tau functions for two-soliton solution as
\begin{eqnarray}
\label{mix1d-59}\hspace{-2cm}\nonumber  f&=&1+E(1,1^*){\rm e}^{\theta_1+\theta^*_1} + E(1,2^*){\rm e}^{\theta_1+\theta^*_2}
+ E(2,1^*){\rm e}^{\theta_2+\theta^*_1} +E(2,2^*){\rm e}^{\theta_2+\theta^*_2}\\
\label{mix1d-60}\hspace{-2cm} &&+E(1,1^*,2,2^*){\rm e}^{\theta_1+\theta_2+\theta^*_1+\theta^*_2},\\
\hspace{-2cm}  g^{(1)}&=&c^{(1)}_1{\rm e}^{\theta_1} + c^{(1)}_2{\rm e}^{\theta_2}
+F(1,2,1^*){\rm e}^{\theta_1+\theta_2+\theta^*_1}+F(1,2,2^*){\rm e}^{\theta_1+\theta_2+\theta^*_2},\\
\label{mix1d-61}\hspace{-2cm}\nonumber  h^{(l)}&=&1+G^{(l)}(1,1^*){\rm e}^{\theta_1+\theta^*_1} + G^{(l)}(1,2^*){\rm e}^{\theta_1+\theta^*_2}
+ G^{(l)}(2,1^*){\rm e}^{\theta_2+\theta^*_1} +G^{(l)}(2,2^*){\rm e}^{\theta_2+\theta^*_2}\\
\hspace{-2cm} &&+G^{(l)}(1,1^*,2,2^*){\rm e}^{\theta_1+\theta_2+\theta^*_1+\theta^*_2},
\end{eqnarray}
where
\begin{eqnarray*}
\hspace{-2cm} && E(i,j^*)=\frac{{\rm i} \sigma_1 c^{(1)*}_i c^{(1)}_j }{ (p_i+p^*_j)^2[p^*_i-p_j + \sum^2_{l=1}\frac{{\rm i}\sigma_{l+1}\rho^2_l}{(p^*_i+{\rm i}\alpha_l)(p_j-{\rm i}\alpha_l)}  ]},\ \
G^{(l)}(i,j^*)=-\frac{p_i-{\rm i}\alpha_l}{p^*_j+{\rm i}\alpha_l} E(i,j^*),\\
\hspace{-2cm} && E(1,1^*,2,2^*)=|p_1-p_2|^2 \Big[\frac{E(1,1^*)E(2,2^*)}{(p_1+p^*_2)(p_2+p^*_1)}
- \frac{E(1,2^*)E(2,1^*)}{(p_1+p^*_1)(p_2+p^*_2)} \Big],\\
\hspace{-2cm} &&G^{(l)}(1,1^*,2,2^*)=\frac{(p_1-{\rm i}\alpha_l)(p_2-{\rm i}\alpha_l)}{(p^*_1+{\rm i}\alpha_l)(p^*_2+{\rm i}\alpha_l)}E(1,1^*,2,2^*),\\
\hspace{-2cm} && F(1,2,i^*)=(p_2-p_1) \Big[ \frac{c^{(1)}_2E(1,i^*)}{p_2+p^*_i}-  \frac{c^{(1)}_1E(2,i^*)}{p_1+p^*_i} \Big].
\end{eqnarray*}

Same as the previous subsection, nonsingular solution requires the denominator $f$ to be real and nonzero.
For this purpose, we rewrite $f$ as
\begin{eqnarray}
\nonumber \hspace{-2cm} f&=&2{\rm e}^{\theta_{1R}+\theta_{2R}}
[{\rm e}^{\frac{\Omega_1+\Omega_2}{2}} {\rm cosh}(\theta_{1R}-\theta_{2R}+\frac{\Omega_1-\Omega_2}{2})
+{\rm e}^{H_{1R}} \cos(\theta_{1I}-\theta_{2I}+H_{1I})\\
\hspace{-2cm}&&+{\rm e}^{\frac{\Omega_3}{2}} {\rm cosh}(\theta_{1R}-\theta_{2R}+\frac{\Omega_3}{2})],
\end{eqnarray}
where ${\rm e}^{\Omega_1}=E(1,1^*),\ \ {\rm e}^{\Omega_2}=E(2,2^*),\ \ {\rm e}^{\Omega_3}=E(1,1^*,2,2^*),\ \
{\rm e}^{H_{1R}+{\rm i}H_{1I}}=E(1,2^*)$.
Thus, one can conclude that $E(i ,i^*)>0,i=1,2$ is a necessary condition to obtain a regular
solution
and ${\rm e}^{\frac{\Omega_1+\Omega_2}{2}}+{\rm e}^{\frac{\Omega_3}{2}}>{\rm e}^{H_{1R}}$ is a sufficient one.

\begin{figure}[!htbp]
\centering
{\includegraphics[height=1.3in,width=6in]{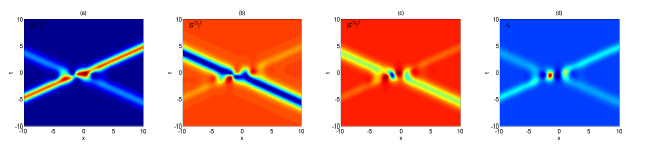}}
\caption{(Color online) Mixed two-soliton solution (one-bright-two-dark soliton for SW components) in the (3+1)-component YO system.
\label{mix1d-fig6}}
\end{figure}

The asymptotic analysis can be performed as in \cite{kanna2013general,kanna2009higher,kanna2014mixed},
whose details are omitted here.
However, it should be remarked here that two-soliton solution for all components in this case
always undertakes elastic collision without shape changing.
This feature is the same as the one of the mixed soliton solution including one-bright-two-dark soliton for SW components in 2D (3+1)-component YO system \cite{kanna2014mixed}.
In Fig. \ref{mix1d-fig6}, we exhibit this mixed-type soliton solution under the same nonlinearity coefficients with Fig. \ref{mix1d-fig5} and
the parameters are given as $\rho_1=\rho_2=\alpha_1=1,\alpha_2=2,
p_1=-\frac{2}{3}-{\rm i}, p_2=\frac{1}{2}+{\rm i},
c^{(1)}_1=1+{\rm i},c^{(1)}_2=2+{\rm i}$ and
$\theta_{10}=\theta_{20}=0$.
Fig. \ref{mix1d-fig6} (a) shows two bright solitons collisions for the SW components $S^{(1)}$, Fig. \ref{mix1d-fig6} (b) and (c) display two dark solitons collisions for the SW components $S^{(2)}$ and $S^{(3)}$ with the different amplitudes, and Fig. \ref{mix1d-fig6} (d) represents two bright solitons collision
for the LW component $-L$.

\begin{figure}[!htbp]
\centering
{\includegraphics[height=1.3in,width=6in]{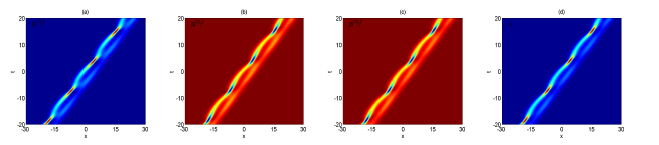}}
\caption{(Color online) Mixed two-soliton bound state (one-bright-two-dark soliton for SW components) in the (3+1)-component YO system.
\label{mix1d-fig7}}
\end{figure}

For the construction of the bound states from the mixed soliton solution (\ref{mix1d-59})-(\ref{mix1d-61}),
one can derive the same condition $p_{1I}=p_{2I}$ as the previous subsection.
By taking the same nonlinearity coefficients in Fig. \ref{mix1d-fig5}, such a bound state with the parameters
$\rho_1=\rho_2=\alpha_1=1,\alpha_2=\frac{3}{2},
p_1=-\frac{1}{2}-\frac{1}{2}{\rm i}, p_2=\frac{9}{10}-\frac{1}{2}{\rm i},
c^{(1)}_1=1,c^{(1)}_2=\frac{1}{3}$ and
$\theta_{10}=\theta_{20}=0$.
is depicted in Fig. \ref{mix1d-fig7}.
Fig. \ref{mix1d-fig7} (a) exhibits two-bright-soliton bound state for the SW components $S^{(1)}$ ,
Fig. \ref{mix1d-fig7} (b) and (c) display two-dark-soliton bound states for the SW component $S^{(3)}$ and $S^{(3)}$ with different amplitudes and two-bright-soliton bound state of the LW component $-L$ is shown in  Fig. \ref{mix1d-fig7} (d).

Here we need to point out that $N$-soliton bound state can be formed in both above mixed soliton solution
for arbitrary combination of the nonlinearity coefficient $\sigma_\ell$ $(\ell=1,2,3)$.
More specifically, if the wave number for the $i$-th soliton among an N-soliton solution is
$p_i=p_{iR}+{\rm i}p_{iI}$ , then $\theta_{iR} = p_{iR} (x + 2p_{iI} t) +\theta_{i0R}$, which suggests
that the velocity for the $i$-th soliton is $-2p_{iI}$.
Thus a multiple bound state in which solitons move with the same
velocity only requires all the $p_{iI}$ to be the same.
However, higher-order soliton bound state cannot exist
in the soliton solution comprised of all dark soliton for the SW components.
As reported in \cite{chen2014multidark},
due to the existence of the parameter constraint in general multi-dark soliton solution,
only two-dark-soliton bound state is possible under the condition that the nonlinearity coefficients $\sigma_\ell$ take different signs.

\section{General mixed soliton solution to the 1D multicomponent YO system}

In the same spirit as the (3+1)-component YO system, the general mixed type soliton
solution to the 1D ($M+1$)-component YO system can be derived by the KP hierarchy reduction method.
It is known that the multi-bright soliton solutions can be
derived from the reduction of the multi-component KP
hierarchy, whereas, the multi-dark soliton solutions are
obtained from the reduction of the single KP hierarchy
but with multiple copies of shifted singular points.
Therefore, if we consider a general mixed soliton solution consisting of $m$ bright solitons and $M-m$ dark
solitons to the SW components in the 1D ($M+1$)-component YO system (\ref{mix1d-01}), we need to start from an $(m + 1)$-component KP hierarchy with $M-m$ copies of shifted singular points in the first component.
Furthermore, by performing the complex conjugation and dimension reductions,
these bilinear equations become ones of multicomponent YO system.
Meanwhile, the general bright soliton solution can be reduced from the tau
functions of the KP hierarchy.
The details are omitted and we only provide the final result here.

To seek for mixed  multi-soliton solution consisting of $m$ bright solitons and $M-m$ dark
ones for SW components, the 1D multicomponent YO system is first converted to the following bilinear form
\begin{equation}
\hspace{-1cm}\begin{array}{l}
\left[ \textmd{i} D_t-D^2_x \right]g^{(k)}\cdot f=0,\ \ k=1,2,\cdots,m,\\
\left[ \textmd{i} (D_t-2\alpha_lD_x)-D^2_x \right]h^{(l)}\cdot f=0,\ \ l=1,2,\cdots,M-m, \\
\left(D_tD_x-2 \sum^{M-m}_{l=1}\sigma_{l+m}\rho^2_l \right)f\cdot f= -2 \sum^m_k\sigma_k g^{(k)}g^{(k)*}-2\sum^{M-m}_{l=1}\sigma_{l+m}\rho^2_l h^{(l)}h^{(l)*},
\end{array}
\end{equation}
through the dependent variable transformations:
\begin{equation}
\begin{array}{l}
 S^{(k)}=\frac{g^{(k)}}{f},\ \
 S^{(l)}=\rho_l\frac{h^{(l)}}{f} {\rm e}^{{\rm i}(\alpha_l x+ \alpha^2_l t)},\ \
 L=-2(\log f)_{xx},\ \
\end{array}
\end{equation}
where $\alpha_l$, $\beta_l$ and $\rho_l(\rho_l>0)$ are arbitrary constants.

Similar to the procedure discussed in Section 2,
one can obtain mixed multi-soliton solution as
follows:
\begin{eqnarray}\label{mix1d-65}
&& f =
\left|
\begin{array}{ccc}
A & I \\
-I & B
\end{array}
\right|,\ \
%%%%%%%%%%%%
h^{(l)} =
\left|
\begin{array}{ccc}
A^{(l)} & I \\
-I & B
\end{array}
\right|,\ \
g^{(k)}=
\left|
\begin{array}{ccc}
A & I & \phi^T \\
-I & B & \textbf{0}^T\\
\textbf{0} & C_k & 0
\end{array}
\right|,
\end{eqnarray}
where $A$, $A^{(k)}$ and $B$ are $N\times N$ matrices whose entries are
\begin{eqnarray*}
&& a_{ij}=\frac{1}{p_i+p^*_j}  {\rm e}^{\theta_i+\theta^*_j},\\
&& a^{(l)}_{ij}=\frac{1}{p_i+p^*_j} \left(-\frac{p_i-{\rm i}\alpha_l}{p^*_j+{\rm i}\alpha_l} \right) {\rm e}^{\theta_i+\theta^*_j},\\
&& b_{ij}=\frac{{\rm i}\sum^m_{k=1} \sigma_k c^{(k)*}_i c^{(k)}_j }{ (p^*_i+p_j)[p^*_i-p_j + \sum^{M-m}_{l=1}\frac{{\rm i}\sigma_{l+m}\rho^2_l}{(p^*_i+{\rm i}\alpha_l)(p_j-{\rm i}\alpha_l)} ]},
\end{eqnarray*}
and $\phi$ and $C_k$ are $N$-component row vectors given by
\begin{eqnarray*}
\hspace{-1cm}&& \phi=({\rm e}^{\theta_1}, {\rm e}^{\theta_2}, \cdots, {\rm e}^{\theta_N} ),\ \
C_k=-(c^{(k)}_1,c^{(k)}_2,\cdots,c^{(k)}_N),\ \
\end{eqnarray*}
with $\theta_i=p_ix-{\rm i}p^2_i t +\theta_{i0}$ and
$p_i$, $\theta_{i0}$ and $c^{(k)}_i$ $(i=1,2,\cdots,N)$ are arbitrary complex parameters.

In the above solution, one necessary condition similar to the (3+1)-component YO system for the existence of an $N$-soliton solution is found as follows
\begin{eqnarray}
\left(\sum^m_{k=1} \sigma_k |c^{(k)}_i|^2\right)\left(2p_{iI} + \sum^{M-m}_{l=1}\frac{\sigma_{l+m}\rho^2_l}{|p_i-{\rm i}\alpha_l|^2} \right)>0,\ \ i=1,2,\cdots,N.
\end{eqnarray}
As reported in \cite{sakkaravarthi2014multicomponent}, the arbitrariness of nonlinearity coefficients $\sigma_\ell$ increases the freedom resulting in rich mixed soliton dynamics.
Here the present solution admits mixed $N$-soliton in the 1D multicomponent YO system for all types of nonlinearity coefficients, including positive, negative and mixed types.

The formula for general mixed soliton solution can be generalized to include all bright and all dark soliton solutions, which is the same as the vector NLS equation \cite{feng2014general}.
More specifically, the general bright soliton solution can be viewed directly as a special case of the bright-dark soliton
one.
When $m=M$, the $N$-bright soliton solution takes the same determinant form as the general bright-dark one.
The general dark soliton solution was also derived from the single KP hierarchy but
the corresponding dimension reduction results in the distinct parameters constraint.
The final expression of the $N$ dark-soliton solution is different from the one of the bright-dark soliton solution.
However, when $m=0$, as discussed in [38], an alternative form of all dark soliton solution
takes the same form as the solution (\ref{mix1d-65}) by redefining the matrix $B$
as an identity matrix ($B_{ij}=\delta_{ij}$)
and imposing the parameters constraint as follows
\begin{eqnarray}
p^*_i-p_i + \sum^{M}_{l=1}\frac{{\rm i}\sigma_{l}\rho^2_l}{|p_i-{\rm i}\alpha_l|^2}=0,\ \ i=1,2,\cdots,N.
\end{eqnarray}
Due to a simple determinant identity shown in \cite{feng2014general},
this form of dark soliton solution coincides with the one in \cite{chen2014multidark}.

\section{Summary and conclusion}

We have constructed the general bright-dark $N$-soliton solution to one-dimensional multicomponent YO system
describing the nonlinear resonant interaction of $M$ short waves with a long wave,
i.e., one-dimensional ($M+1$)-component YO system.
This solution exists in the original system for all possible combinations of nonlinearity coefficients including positive, negative and mixed types.
Taking the (3+1)-component YO system as an example, we have deduced two kinds of the general mixed $N$-soliton solution (two-bright-one-dark soliton and one-bright-two-dark one for SW components) in the form of Gram determinant
by using the KP hierarchy reduction method.
Then, the same analysis was extended to obtain the general mixed solution consisting of $m$ bright solitons and $M-m$ dark
ones for SW components in the ($M+1$)-component YO system.
The expression of the mixed solution also contains the general bright and dark $N$-soliton solution.

For the dynamics of the mixed solitons, in parallel with the 2D multicomponent YO system \cite{kanna2014mixed},
the energy exchanging collision of mixed solitons can be realized only in the bright parts of the mixed solitons
and is possible only if the bright parts of the mixed solitons appear at least in
two SW components.
Such an interesting phenomenon can also be found in 1D (3+1)-component YO system as discussed in previous section, as well as in the 1D multi-component YO system.
In addition, the related analysis regarding mixed-soliton bound state in 1D (3+1)-component YO system implies that arbitrary higher-order mixed-soliton bound state in 1D multi-component YO system can also be formed.

\section*{Acknowledgments}
J.C. appreciates the support by the China Scholarship
Council. The project is supported by the Global Change
Research Program of China (No. 2015CB953904), National
Natural Science Foundation of China (Grant Nos. 11275072, 11435005 and 11428102), Research Fund for the Doctoral
Program of Higher Education of China (No.
20120076110024), The Network Information Physics Calculation
of basic research innovation research group of China
(Grant No. 61321064), Shanghai Collaborative Innovation
Center of Trustworthy Software for Internet of Things (Grant
No. ZF1213), Shanghai Minhang District talents of high level
scientific research project and CREST, JST.

\end{document}